\documentclass{article}
\usepackage{graphics}
\newcommand{\beq}{\begin{equation}}
\newcommand{\eeq}{\end{equation}}
\newcommand{\bdis}{\begin{displaymath}}
\newcommand{\edis}{\end{displaymath}}
\newcommand{\bea}{\begin{eqnarray}}
\newcommand{\eea}{\end{eqnarray}}
\newcommand{\barr}{\begin{array}}
\newcommand{\earr}{\end{array}}
\begin{document}
 
\title{Effect of impact energy on the shape of granular heaps}
 
\author{Yan Grasselli$^{1,2}$, Hans J. Herrmann$^{1,2}$, Gadi Oron$^2$  
and Stefano Zapperi$^{2}$\\\small
$^1$ICA 1, University of Stuttgart, 
Pfaffenwaldring 27, 70569 Stuttgart, Germany\\\small
$^2$PMMH-ESPCI, 10 Rue Vauquelin, 75231 Paris CEDEX 05, France.}
\date{\today}
\maketitle

\begin{abstract}
We study experimentally the shape of a granular heap formed pouring 
a granular material into a vertical Hele-Shaw cell and
analyze the effect of the grain impact energy. 
We propose a continuous model 
for the steady profile of the heap that explicitly considers  
energy dissipation of flowing grains through inelastic collisions. 
We solve the model analytically and analyze the resulting height
profile as a function of several parameters,
such as the restitution coefficient of the grains and their
impact energies. We find good agreement between theory 
and experiments.
\end{abstract}
 

\section{Introduction}

The study of granular materials has received  wide 
attention in the last few years \cite{RSPS,WIL,HW}, because 
of the challenging problems posed by this
state of matter. Sand heaps are among 
the most familiar granular systems that have been
studied experimentally and theoretically. Despite the apparent
simplicity a sand heap presents a number of
intriguing phenomena, such as avalanches \cite{nagel} 
internal texture \cite{witt},
segregation and stratification \cite{makse}. 
Several theoretical models have been employed to describe
the dynamics and the static profile of a sandpile, ranging
from purely numerical models, such as lattice gas
\cite{KK,PH}
and molecular dynamics simulations \cite{mat}, to discrete lattice models
\cite{hovi} and continuous equations \cite{bcre,dg1}. 

Granular heaps can be generated by slowly pouring grains
(i.e. spherical beads, sand) from a given height over 
a rigid support (3D piles) or into a vertical Hele-Shaw cell (2D piles).
The pile typically displays a well defined slope, corresponding
to the angle of repose, with clear deviations at the top and
at the tail. In particular, the tail of a 2D pile was found experimentally
to follow a logarithmic dependence \cite{alonso,grass}. 
This result was anticipated
by the solution of a continuous model solved in a 2D silo geometry
\cite{dg1} and by theoretical arguments \cite{alonso}. 
The angle of repose depends on the characteristics of the material
such as density, humidity, packing history and boundary conditions.

Here we analyze the shape of the top of the granular heap,
focusing on the effect of the impact energy of the particles.
To this end, we perform a set of experiments with glass and leads beads
poured at very slow rate into a vertical Hele-Shaw cell and study
the changes of the heap profile when grains are poured
from different heights. 
We propose a simple continuous model to describe the
steady-state properties of the grains flowing on the top
of the pile, providing an explicit expression for the 
heap profile. The model treats explicitly the dissipation
of the kinetic energy of the flowing grains due to inelastic
collisions and provides a good fit to the experimental data
on glass and lead beads.

\section{Experiments}

Experiments are performed within a vertical Hele Shaw cell 
(20x30cm) with a fixed thickness of
$5 mm$ \cite{note}. Lead beads, with diameter $d=0.2cm$ 
are poured through a funnel placed above the center of the cell. 
We carefully determine the impact energy of the particles , 
controlling the distance $Z_0$ between the top of the heap and the
funnel: the impact energy is thus $e_0 = mgZ_0$.
The falling heights range from $Z_0=1cm$ to $Z_0=14 cm$
and are determined with a with a precision of $\Delta Z_0 = 0.5 cm$.
Experiments are performed at very low particle
flux, dropping one particle at a time through the funnel.
In the range of falling heights employed \cite{crater}, 
the reorganization of the pile after impact 
is usually very weak and the particle jumps
down the slope and is eventually trapped, giving rise to a local 
change in the slope of the pile. The dynamics of the beads can
be sometimes very complex: we observe collisions with the walls
of the cell, rolling motion, ejection of other particles and
occasionally big avalanches of a few layers.

\begin{figure*}
\resizebox{\textwidth}{!}{\rotatebox{-90}{
\includegraphics{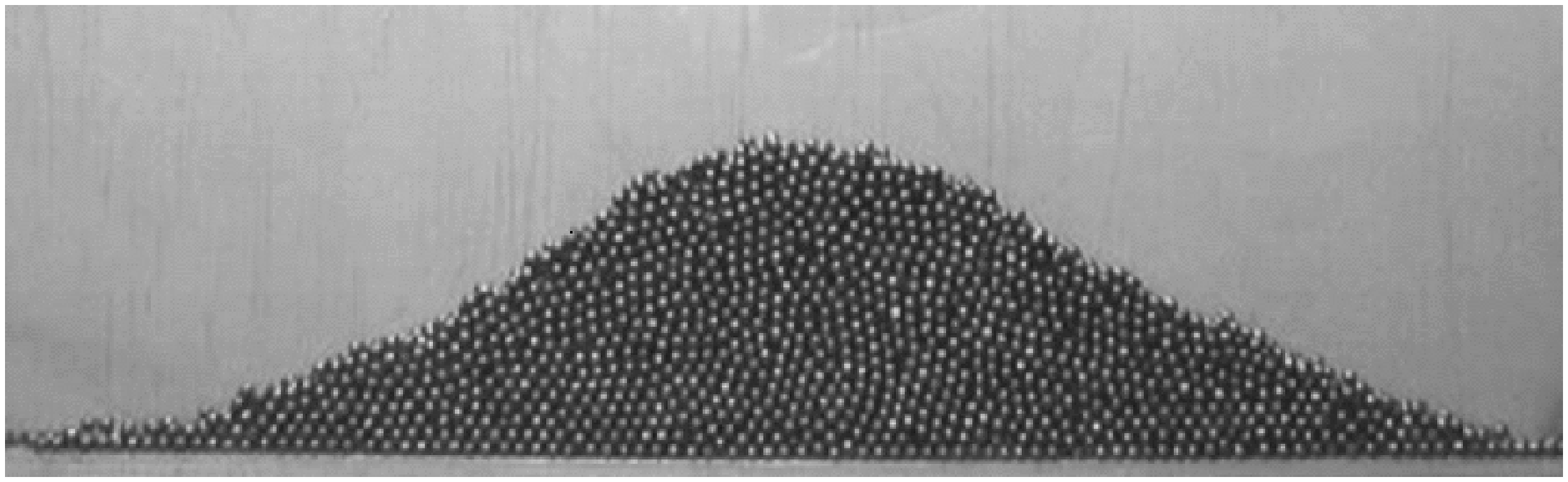}}}
\caption{An example of a heap obtained
experimentally with lead beads.}
\label{fig0}
\end{figure*}

\begin{figure}
\resizebox{0.75\hsize}{!}{%
  \includegraphics{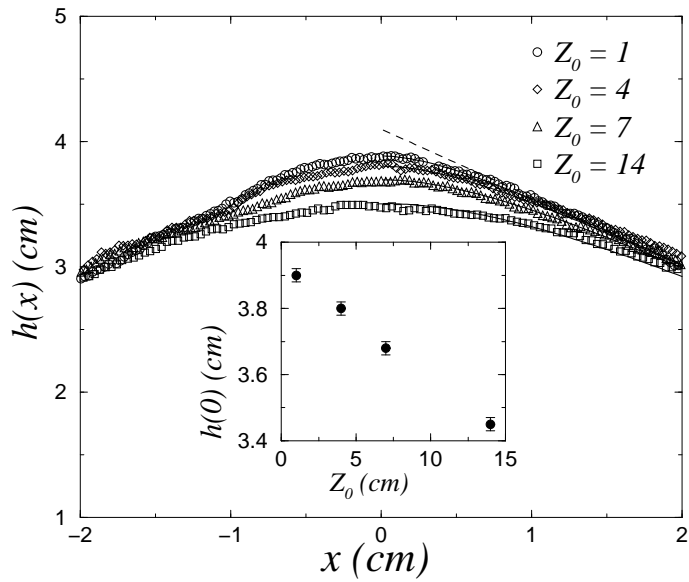}
        }
\caption{The height profiles of the heaps obtained
experimentally with lead beads 
for different impact energies $Z_0$ fitted with 
Eq.~\protect\ref{eq:hexp}. The dashed line has a slope
$\theta_c\simeq 0.54$. In the inset we show
the linear behavior of $h(0)$ with $Z_0$. }
\label{fig1}
\end{figure}

\begin{figure}
\resizebox{0.75\hsize}{!}{%
  \includegraphics{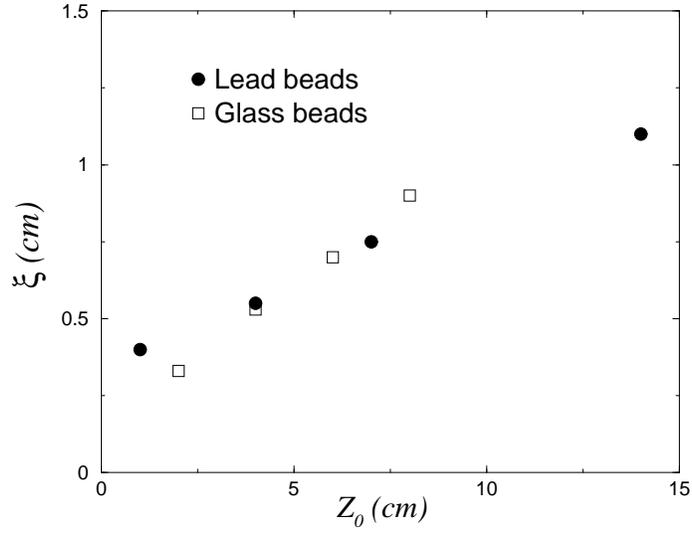}
        }
\caption{
Linear relation between 
the characteristic lenght $\xi$ and the impact energy $Z_0$
for lead and glass beads.}
\label{fig2}
\end{figure}

After a short transient stage, the heap grows steadily and the profile
translates at constant velocity, proportional to the input flux.
In this regime we retrieve the experimental heap profiles (see
Fig.~\ref{fig0})from image analysis and we average them. 
The averaging is done superimposing the profiles
obtained at different time steps in the steady state.
In Fig.~\ref{fig1} we show the averaged profiles for different impact energies.
It can be observed that the heap has a flat top and shows 
a well defined angle of repose, which turns out to be essentially
independent on $Z_0$, at least for the experimental values used.
When the tails of the heaps obtained with different $Z_0$ 
are superimposed, we observe that the maximum height $h(0)$ decreases
linearly with $Z_0$ (inset of Fig.~\ref{fig1}). 
We find that the shape of the heap is well 
described by the form 
\beq
h(x) = A-\theta_c x -\theta_c\xi e^{(-x/\xi)},
\label{eq:hexp}
\eeq
where $\theta_c$ is the angle of repose, $\xi$ is 
a correlation length and $A$ is an inessential constant, that can be
eliminated through an appropriate vertical translation of the profile.
From Eq.~(\ref{eq:hexp}) we have that $h(0)=A-\theta_c\xi$, which
implies that $\xi$ is proportional to $Z_0$, which is verified
by the fitting parameters (see Fig.~\ref{fig2})
This result reflects the fact that $Z_0$ is, together with the particle
diameter, the only characteristic length present in the system.

\begin{figure}
\resizebox{0.75\hsize}{!}{%
  \includegraphics{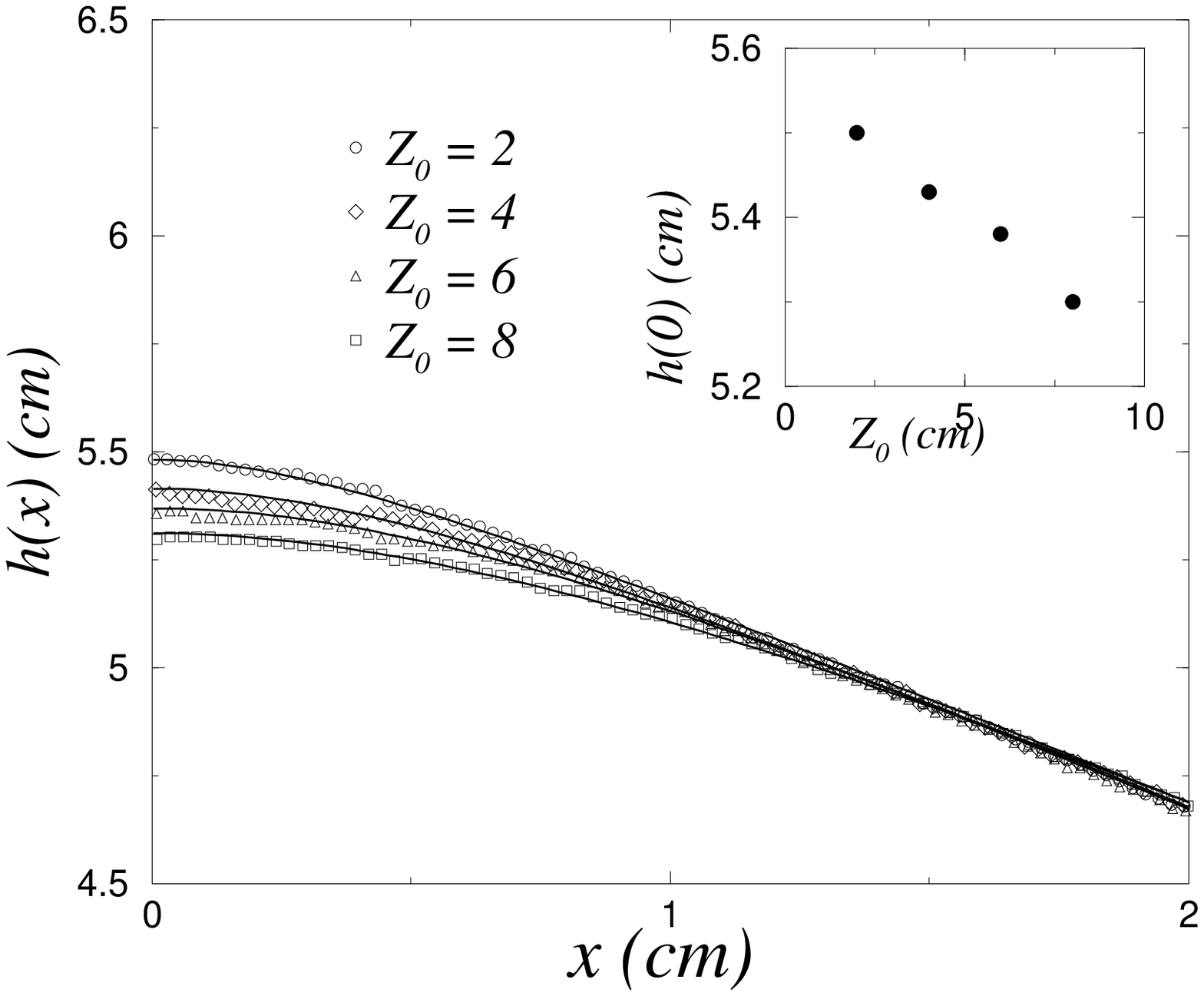}
}


\caption{The height profile obtained in the experiment with
glass beads, for different impact energies $Z_0$ at the lowest
available input flux $W=0.32 g/s$. The lines are the fit with 
Eq.~\protect\ref{eq:hexp}.
In the inset we show the linear dependence of $h(0)$ with $Z_0$.}
\label{fig3}
\end{figure}

The basic features of these results are quite robust and do not depend
on the particular material used. We verified this by forming heaps
with glass beads of considerably smaller 
diameter ($d=0.025cm$), poured at low
flux $W = 0.32 g/s$.  In this case $Z_0$ can be controlled very 
precisely by moving the outlet up at the same 
velocity as the growth of the top of the heap.
In this experiment, grains are poured from one of the
two sides of the Hele-Shaw cell.
We see in Fig.~\ref{fig3} that also in this case $h(0)$ is linear in $Z_0$
and the decay is reasonably well fitted by an exponential,
excluding from the fitting region the first 0.5cm, 
which correspond to the radius of the funnel. 
Finally in this case, the measured angle of repose displays a small
dependence on $Z_0$, of the order of $\Delta\theta= 0.05$.
Looking at Fig.~\ref{fig2}, we note that the linear 
dependence of $\xi$  from $Z_0$ is steeper 
in the case of glass beads by roughly a 
factor $2$. The origin of this effect is probably due to the 
different restitution coefficients of the materials.

\section{Model}

In order to understand the experimental observations,
we consider a sandpile model in which energy dissipation
is explicitly treated. The complete dynamical evolution of
all the grains and their energy dissipation is extremely difficult
to treat explicitly. The evolution of the heap is indeed
a very complex phenomenon, being due to the way momentum is transfered
during particle collisions, frictional properties and other
forms of energy dissipation. However, we note that in our experiment
the growth of the heap displays a remarkable stationary state, which
we can try to describe with a simplified model taking 
into account essentially only energy dissipation.
   
Alonso et al. \cite{hovi} have introduced
a simple automaton model which has been used to derive the angle of
repose of a pile in terms of microscopic parameters, such
as the restitution coefficient of the grains. 
In this model, grains are poured one by one
on the top of a heap with an initial kinetic energy $e_0$.
As the grains fall down the pile, parameterized by the local height $h(x)$, 
their kinetic energy $e$ decreases due to inelastic collisions
and increases because of gravity as
\beq
e(x+\delta)=r(e(x)+mg (h(x)-h(x+\delta))),
\label{eq:endisc}
\eeq
where $m$ is the mass of the grains, $\delta$ 
is the characteristic size of the
jumps, $g$ is the gravity acceleration 
and $r$ is an ``effective'' restitution coefficient. 
The grains jump to the right 
until their kinetic energy is less than a threshold $u$. When a grain
comes to rest at $x$ the height $h$ of the column $x$ is increased by one unit
and a new grain is poured at the origin. This model has been simulated
and, despite its simplicity, was shown 
to explain several experimental results \cite{rimmele}.

We study here a continuous version of the model in which
the steady-state kinetic energy of the rolling grains
follows Eq.~(\ref{eq:endisc})
\beq
\delta\frac{de(x)}{dx} = (r-1)e(x)+r\gamma(x),
\label{eq:encont}
\eeq
where $\gamma(x)\equiv-mg\delta dh/dx$ measures the local slope. 
Eq.~(\ref{eq:encont}) describes the steady-state energy profile
in terms of a  given slope profile $\gamma(x)$. 
To close the problem, we need  to specify how the steady-state
slope profile depends on the energy profile. We assume that
the system in the steady state is at the verge of stability and
that the variations of the local slope compensate
the difference between the kinetic energy $e$ and the trapping
energy $u$: the slope will decrease for $e>u$ and increase
for $e<u$. The simplest equation of this form is given by
\beq
\delta\frac{d\gamma(x)}{dx} = \Gamma (u -e(x)),
\label{eq:gamma}
\eeq
where $\Gamma$ is a phenomenological parameter. 
Eq.~(\ref{eq:gamma}) implies that for large $x$ the
grains are at the threshold of mobility $e(x) \simeq u$.

Finally, we impose the boundary conditions $e(0)=e_0=mg Z_0$ and 
$\gamma(0)=0$. We note that by dividing each term of the equation by $mg$ 
all the quantities can be expressed in terms of lengths.
Eqs.~(\ref{eq:encont}-\ref{eq:gamma}) are linear and can be
solved in order to obtain an explicit expression for
the profile of the pile. It is convenient to transform the pair of equations
into an homogeneous one using the transformation 
\beq
E(x)= e(x)-u,~~~~~~G(x) = \gamma(x)+ (1-r)u/r.
\eeq
The solution of the homogeneous equations
\bea
\delta\frac{de(x)}{dx} = (r-1)E(x)+rG(x), \\
\delta\frac{d G(x)}{dx} = -\Gamma E(x),
\eea
is given by 
\bea
e(x)=u+E_{+}e^{\lambda_+x/\delta}+E_{-}e^{\lambda_{-}x/\delta}\\
\gamma(x)=(r-1)u/r+G_{+}e^{\lambda_+x/\delta}+G_{-}e^{\lambda_{-}x/\delta}
\eea
where $\lambda_{\pm}=(r-1\pm\sqrt{(r-1)^2-4r\Gamma})/2$ 
are the eigenvalues of the matrix
\beq
\cal{L}\equiv \left(
\begin{array}{cc}
r-1 & r \\
-\Gamma & 0\\
\end{array}\right)
\eeq
The coefficients $E_{\pm}$ and $G_\pm$ can be explicitly obtained
imposing the boundary conditions.

In the limit of large $x$ the slope of the pile is given by
$\gamma_\infty = (r-1)u/r$, from which we obtain 
the angle of repose $\theta_c=(1-r)u/(mg\delta r)$. 
Since the real part of $\lambda_{\pm}$ is
negative for all the values of the parameters the model
has a well defined angle of repose. 
The relaxation towards the angle of repose is expressed by
the sum of two exponentials with characteristic lengths
$\xi_{\pm} \equiv \delta/\lambda_{\pm}$. Since $\xi_{+} > \xi_{-}$,
the relaxation is dominated by $\xi_{+}$ that we can identify 
with $\xi$ of Eq.~\ref{eq:hexp}.
Finally, integrating the equation for $\gamma(x)$,
we obtain an analytical form for the shape of the pile that
can be compared with experiments
\[
h(x)=h(0)-(1-r)ux/(mg\delta r)-
\]\beq
\delta[G_+/\lambda_+(
e^{\lambda_+ x/\delta}-1)+G_-\lambda_-(e^{\lambda_- x/\delta}-1)].
\eeq
The exponential relaxation to the angle of repose is 
experimentally observed, although in the present form
the model can not account for the impact energy dependence
of the relaxation length.

An original assumption of Eq.~(\ref{eq:endisc}) was that
the jump length $\delta$ is constant during the grain motion.
The problem of a ball rolling and jumping over a rough
inclined plane has received a large attention in recent years
\cite{ball1,ball2}. It has been experimentally observed that as
a function of the angle of the inclined plane the motion is
decelerated or accelerated, with a region in between where the
motion is effectively viscous. The energy relaxation was recently
studied in Ref.~\cite{ball2}, where the stopping length in 
the decelerated regime was measured as a function of the initial
energy. The result clearly indicates that
the stopping length is linearly dependent on the initial energy.
We can show this fact by a simple argument.
Consider an inelastic ball hitting a plane with energy $e_0$
and let $\alpha$ be the angle between the plane and the 
particle velocity before the impact.
We consider for simplicity that normal and tangential restitution
coefficients are both equal to $r$ so that at each jump $n$ the energy
decreases by $e_n=r e_{n-1}$. 
A straightforward calculation shows that
the jump length decreases with $n$ as $\delta_n=2e_0 r^n \sin(2\alpha)/(mg)$,
which yields $e(x)=e_0 [1-mgx(1-r)/(2e_0 \sin(2\alpha))]=e_0(1-x/\xi)$.  
We see here that the characteristic length for energy relaxation $\xi$
is linear in $e_0$, in agreement with the experimental result
seen in Fig.~\ref{fig2}

\section{Conclusions}

In conclusion, we have studied the effect of impact energies
on the shape of the sand heap. Experiments show an exponential
relaxation of the slope to the angle of repose, with a characteristic
length that is linearly dependent on the impact energy of the grains.
We model this process by simple relaxation equations and
explain theoretically the experimental results.

\noindent
S. Z. is supported by EC TMR Research Network contract ERBFMRXCT960062.

\newpage

\begin{thebibliography}{99}
\bibitem{note}
  The tail of the
  heap is affected by the presence of a confining 
  geometry (i.e. silos) \cite{dg1,grass}. This is not the case 
  for the top of the heap, which we are studying here.
\bibitem{crater}
	We are currently performing experiments with large
	impact energies, with $Z_0$ of the order of a few meters.
	In this case the behavior is completely different from the
	one studied here: we observe a large reorganization
	of the pile after each impact and the formation of a crater
	at the top of the pile.
\end{thebibliography}
\end{document}